\documentstyle[aps,prl,preprint,floats,epsfig]{revtex}

\begin{document}

\preprint{\vbox{\hbox{CLNS 01-1772}    
                \hbox{CLEO 01-25} 
                }}  

\vskip 0.5cm

\title{ Observation of Exclusive 
           $\bar{B}$ $\to$ $D^{(*)} K^{*-}$ Decays }

\author{CLEO Collaboration}
\vskip 1.0cm 
\date{December 11, 2001}
\maketitle 

\tighten

\begin{abstract} 
We report the first observation of the
 exclusive decays $\bar B\to D^{(*)}K^{*-}$,
using 9.66 $\times$ 10$^{6}$ $B\overline{B}$ pairs collected 
at the $\Upsilon(4S)$ with the CLEO  detector. 
We measure the following branching fractions: 
${\cal B}(B^{-} \to D^{0} K^{*-} ) =  (6.1~\pm~1.6~\pm~1.7)\times 10^{-4},$
${\cal B}(\bar{B^{0}}\to D^{+}K^{*-}) =  (3.7~\pm~1.5~\pm~1.0)\times 10^{-4},$
${\cal B}(\bar{B^{0}}\to D^{*+}K^{*-}) = (3.8~\pm~1.3~\pm~0.8)\times 10^{-4}$
and 
${\cal B}(B^{-} \to D^{*0} K^{*-} ) =  (7.7~\pm~2.2~\pm~2.6)\times 10^{-4}.$
The $\bar B$ $\to$ $D^{*}K^{*-}$ branching ratios are 
the averages of those corresponding to the
00 and 11 helicity states. 
The errors shown are statistical and systematic, respectively. 
\end{abstract} 


\newpage
{
\renewcommand{\thefootnote}{\fnsymbol{footnote}}

\begin{center}
R.~Mahapatra,$^{1}$ H.~N.~Nelson,$^{1}$
R.~A.~Briere,$^{2}$ G.~P.~Chen,$^{2}$ T.~Ferguson,$^{2}$
G.~Tatishvili,$^{2}$ H.~Vogel,$^{2}$
N.~E.~Adam,$^{3}$ J.~P.~Alexander,$^{3}$ C.~Bebek,$^{3}$
% ...
\end{center}
 
\small
\begin{center}
$^{1}${University of California, Santa Barbara, California 93106}\\
$^{2}${Carnegie Mellon University, Pittsburgh, Pennsylvania 15213}\\
$^{3}${Cornell University, Ithaca, New York 14853}\\
% ...
$^{6}${University of Illinois, Urbana-Champaign, Illinois 61801}\\
% ...
\end{center}
 
}
\newpage 


The study of $CP$  violation in $B$ mesons
provides a decisive test of the %
$CP$  violation mechanism 
in the Standard Model (SM).
In the SM, $CP$ violation is the consequence
of the complex phase in the Cabibbo-Kobayashi-Maskawa (CKM)
quark-mixing matrix~\cite{ckm}.
Comprehensive tests of the SM predictions on $CP$ violation 
require  precision measurements of the three sides and three angles 
of the CKM unitary triangle~\cite{babarbook}. 
The angle $\gamma$ can be 
measured using $\bar B$ $\to$ $D^{(*)} K^{(*)}$ 
decays~\cite{babarbook,gronau}.
The decay $B^{-}$ $\to$ $D^{0} K^{-}$ was first observed at CLEO~\cite{cleodk}
and confirmed by BELLE \cite{BELLE}. 
In this Letter, we report the first observation of the exclusive
hadronic $B$ decays $\bar{B}$ $\to$ $D^{(*)} K^{*-}$. 
Charge-conjugate modes are implied throughout this Letter. 


The data were collected with two configurations 
(CLEO II~\cite{CLEOII} and CLEO II.V~\cite{CLEOII.V}) of the
CLEO detector at the Cornell Electron Storage Ring (CESR). 
The data consist of 9.13 fb$^{-1}$ taken at the $\Upsilon$(4S), which 
corresponds to $9.66 \times 10^6$ $\mbox{$B\bar{B}$}$ pairs, 
and 4.35 fb$^{-1}$ taken below $B\overline{B}$ threshold, which is
used for continuum background studies. We assume that 
the produced $B^+B^-$ rate is the same as $B^0\bar B^0$~\cite{bbratio}
at the $\Upsilon(4S)$.


Signal $B$ meson candidates are fully reconstructed by
combining detected photons and charged pions and kaons. 
The detector elements most important for the results presented 
here are the tracking system, which consists of several  
concentric detectors operating inside a 1.5 T superconducting solenoid, 
and the electromagnetic calorimeter, consisting of 7800 
CsI(Tl) crystals. For CLEO II, the tracking system consisted of a 6-layer
straw tube chamber, a 10-layer precision drift chamber, and a
51-layer main drift chamber. The main drift chamber also provided a
measurement of the specific ionization loss, $dE/dx$, used for
particle identification.  For CLEO II.V, the straw tube
chamber was replaced by a 3-layer, double-sided silicon vertex
detector, and the gas in the main drift chamber was changed from 
an argon-ethane to a helium-propane mixture.

The particles in the final state are identified via the decay modes 
$K^{*-} \to K^{0}_{S}\pi^{-}$ with 
$K^{0}_{S} \to \pi^{+}\pi^{-}$;
$D^{0}$  $\to$ $K^{-}\pi^{+}$, $K^{-}\pi^{+}\pi^{0}$ and
               $K^{-}\pi^{+}\pi^{+}\pi^{-}$;
$D^{+}$  $\to$ $K^{-}\pi^{+}\pi^{+}$;
$D^{*+}$ $\to$ $D^{0}\pi^{+}_{s}$;
$D^{*0}$ $\to$ $D^{0}\pi^{0}$ and $D^{0}\gamma$.
Reconstructed tracks are required to pass quality cuts based
on the track fit residuals,  the impact parameter in both the
$r$--$\phi$ and $r$--$z$ planes and the number of drift chamber
measurements. 
The $dE/dx$ measured by the main drift chamber
is used to distinguish kaons from pions. 
Electrons are selected based on $dE/dx$ information and the ratio of the 
associated shower energy in the calorimeter to the measured track momentum. 
Muons are selected  by requiring that charged tracks penetrate 
more than five interaction lengths of material. Any hadron candidate 
is required not to be identified as an electron or a muon. 
Pairs of charged tracks used to reconstruct the $K^{0}_{S}$'s  
(via $K^{0}_{S} \rightarrow \pi^+ \pi^-$) are required to have a common 
vertex displaced from the primary interaction point. The invariant 
mass of the two charged pions is required to be within 3 standard 
deviations ($\sigma\approx3~$ MeV) of the nominal $K^{0}_{S}$ mass~\cite{pdg}.
Furthermore, the $K^{0}_{S}$ momentum vector, obtained from
a kinematic fit of the charged pions' momenta, is
required to point back to the beam spot.
To form $\pi^0$ candidates, pairs of photon candidates with an invariant mass 
within $[-3.0,2.5]\sigma$ ($\sigma\approx 6$ MeV) 
of the nominal $\pi^0$ mass are kinematically fitted 
with the mass constrained to the nominal $\pi^0$ mass~\cite{pdg}.
The soft photon from $D^{*0} \to D^{0}\gamma$ decay is required to have 
an energy
of 100 MeV or greater, 
and a $\pi^0$ veto within $[-4.5,3.5]\sigma$ of the $\pi^0$ mass is applied. 

  $B$ meson candidates are identified 
through their measured mass and energy. 
There are two key variables for full reconstruction of $B$ mesons at CLEO,
which take advantage of the fact that the $B$ meson energy is the same as the
beam energy.
One is the beam-constrained mass of 
the candidate which is defined as 
$M_{B} \equiv \sqrt{E_{\rm beam}^2 - |{\bf p}|^2}$, 
where $\bf p$ is the measured momentum of the candidate and $E_{\rm beam}$ is
the beam energy. 
The second observable, $\Delta E$, is defined to be the sum of the
energies of the decay products of the $B$ candidate minus the 
beam energy, {\it i.e.},  
$\Delta E \equiv  E_{B}$ $-$ $E_{\rm beam}$.
$|\Delta E| $ will be large  if a decay product
 of the $B$ candidate has been lost or
assigned the wrong particle species. 
For fully-reconstructed $B$ meson decays, the $M_B$ distribution peaks at 
5.28 GeV with resolution around 3 MeV, and $\Delta E $ peaks at 0 GeV, 
 with 
a resolution of about 15 MeV.

The $K^{*-}$ is required to have a mass within 75 MeV  
of its nominal mass~\cite{pdg} and  
$D^{0}$  $\to$ $K^{-}\pi^{+}$, $K^{-}\pi^{+}\pi^{0}$,
              $K^{-}\pi^{+}\pi^{+}\pi^{-}$ 
candidates are required to have masses within 
 16 MeV, 25 MeV and 14 MeV ($2\sigma$)  
of their nominal masses respectively~\cite{pdg}.  
For $D^{0}\to K^{-}\pi^{+}\pi^{0}$, we further impose a Dalitz 
weight cut \cite{Dalitz} which reduces the 
background by about 70\% with only about a 20\% efficiency loss.   
The mass differences between $D^{*+}(D^{*0})$ and $D^{0}$ are
required to be within 2.0$\sigma$ of their nominal values~\cite{pdg}
($\sigma\approx0.8$ MeV for $D^{*+}\to D^0\pi^+_{s}$, 1.0 MeV 
for $D^{*0}\to D^0\pi^0$ and 
4.0 MeV for $D^{*0}\to D^0\gamma$).
Because the $K^{*-}$ from $\bar B\to D K^{*-}$ decays is polarized, 
 we further require
$|\cos\alpha|$ $>$ 0.4, where $\alpha$ is the helicity angle 
between one of the $K^{*-}$ decay products in the $K^{*-}$ 
rest frame and the direction of the $K^{*-}$ momentum
in the $B$ rest frame.

The main background comes from continuum $e^{+}e^{-}$ $\to$ 
$q\bar q$ events, where $q$ $=$ $u$, $d$, $s$, $c$. 
To suppress this background, we require that the ratio 
$R_2$ of the second to the 
zeroth Fox-Wolfram moments, determined using charged tracks and
unmatched neutral showers, be less than 0.5.
To further reduce continuum background, we require that 
the absolute value of the cosine
of the angle between the sphericity axis~\cite{sphericity} 
of the candidate tracks and the sphericity axis of the rest of the event
 be less than 0.9.
The distribution of the cosine of this angle should be flat for 
$B$ mesons and strongly peaked near $\pm$1.0 for the continuum background. 

Each $K^{*-}$ candidate is combined with any $D^{(*)}$ candidate 
in the event to form a $\bar{B} \to D^{(*)} K^{*-}$ candidate. 
All the $\bar{B} \to D^{(*)} K^{*-}$ candidates are required to have
$\Delta E$ within 2.5$\sigma$ of zero,  with 
the resolution varying from 12 to 20 MeV depending on the decay mode.  
In case of an event with multiple candidates, we select the one 
with the smallest $|\Delta E|$. 
Studies show that no significant bias is introduced by this procedure.

\begin{figure} [t]
\epsfxsize=0.40\textwidth
\center{\mbox{\epsffile{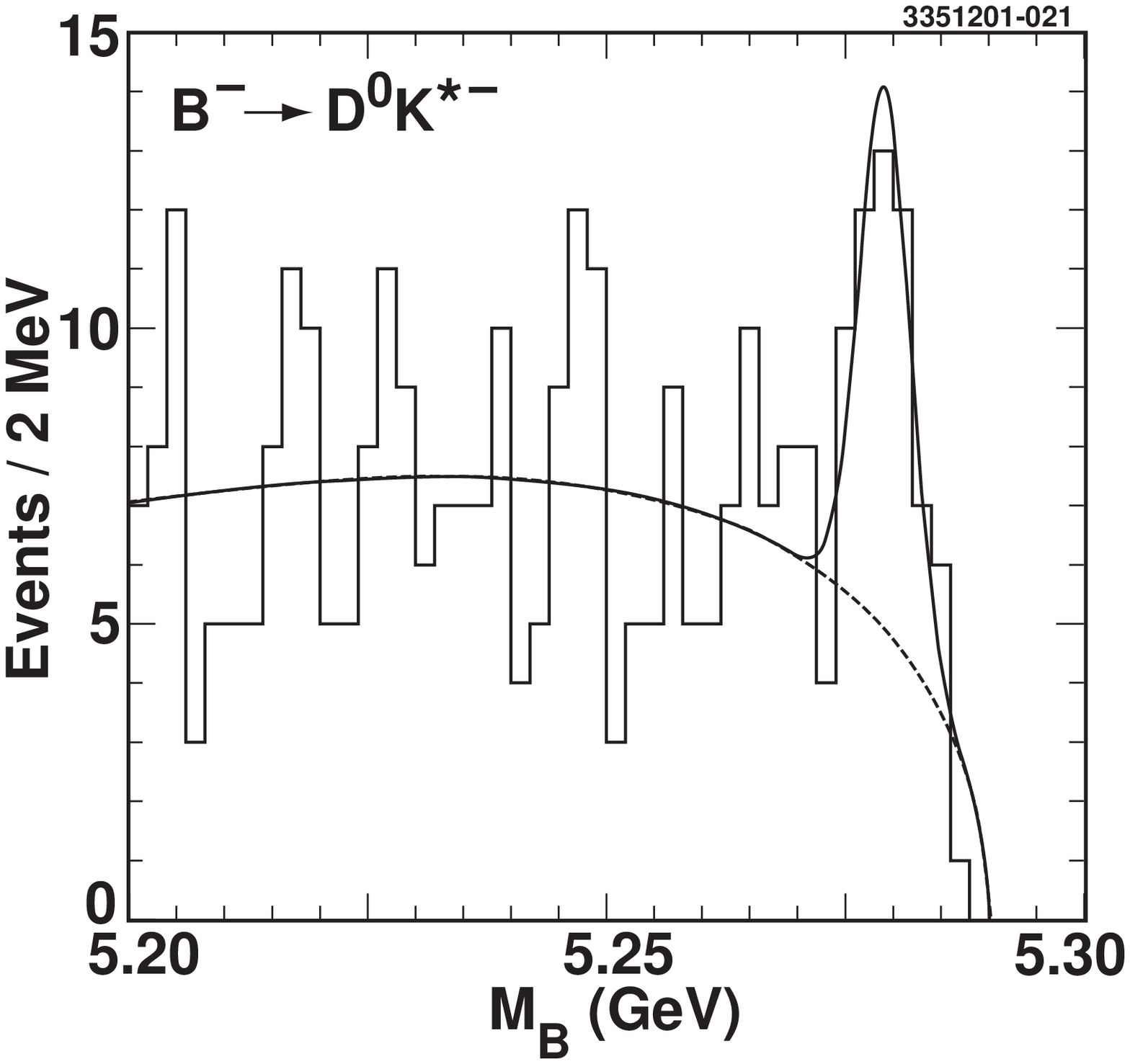} 
       \epsfxsize=0.40\textwidth\epsffile{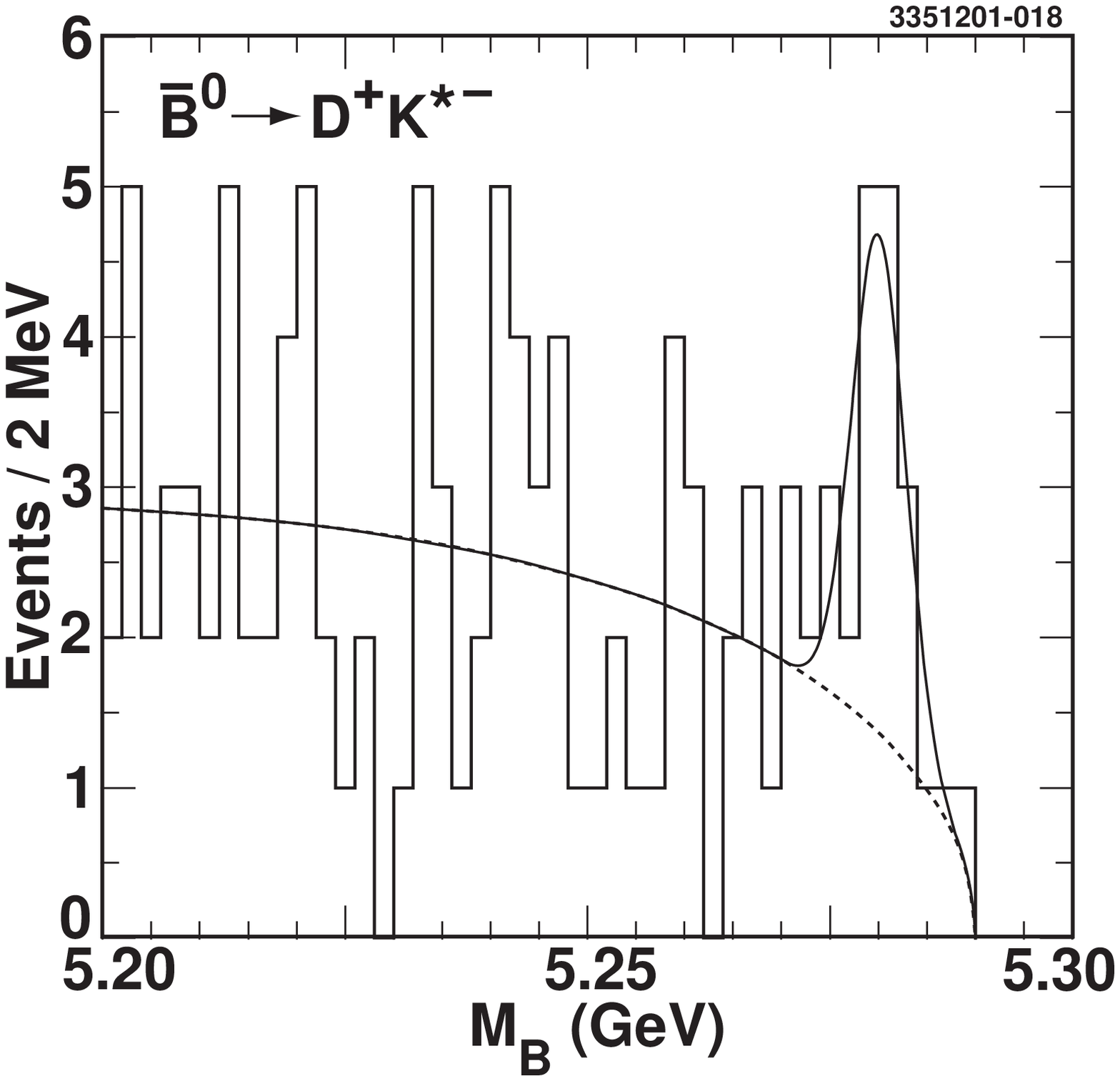}}}
\epsfxsize=0.40\textwidth
\center{\mbox{\epsffile{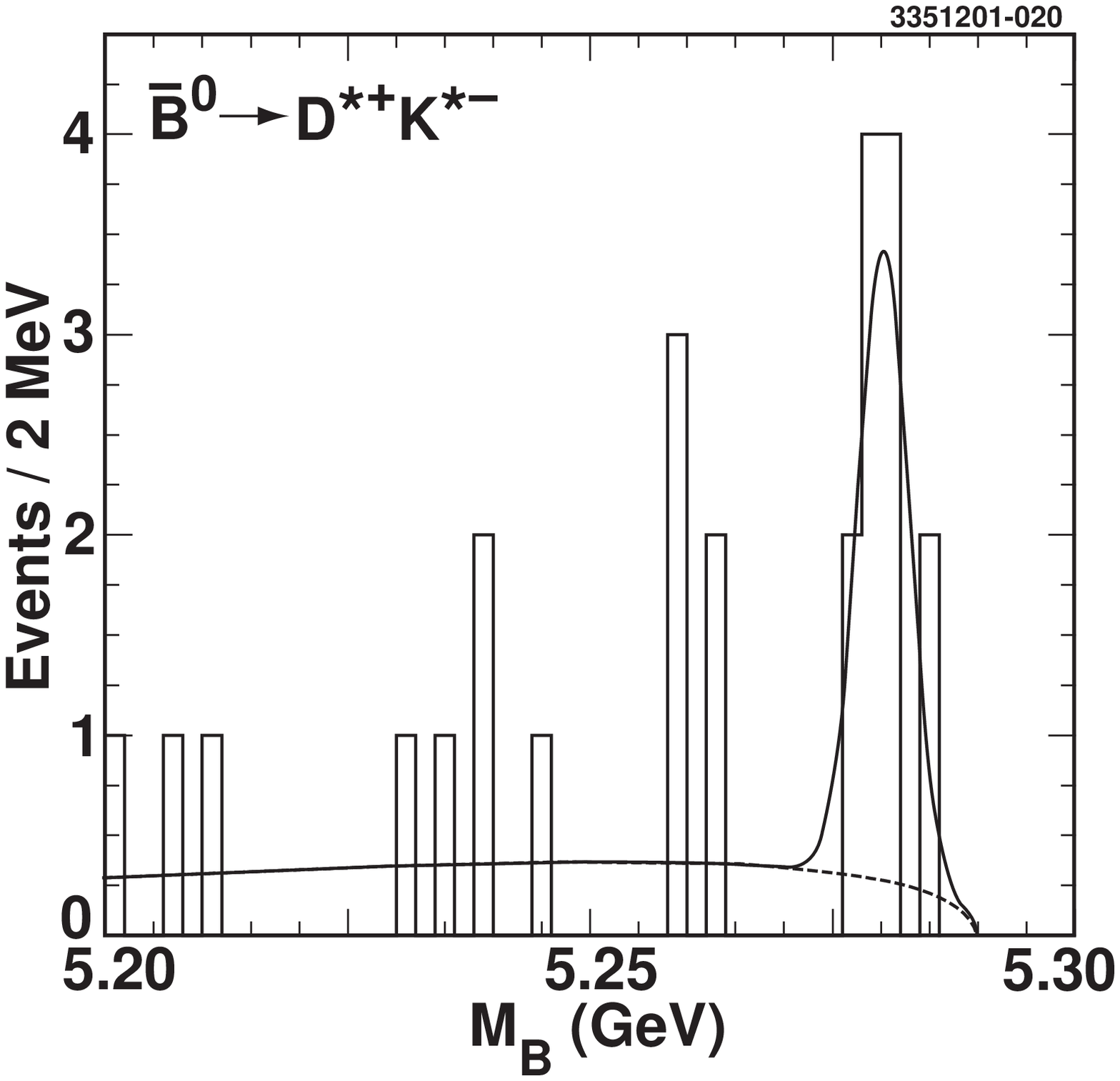}
      \epsfxsize=0.40\textwidth\epsffile{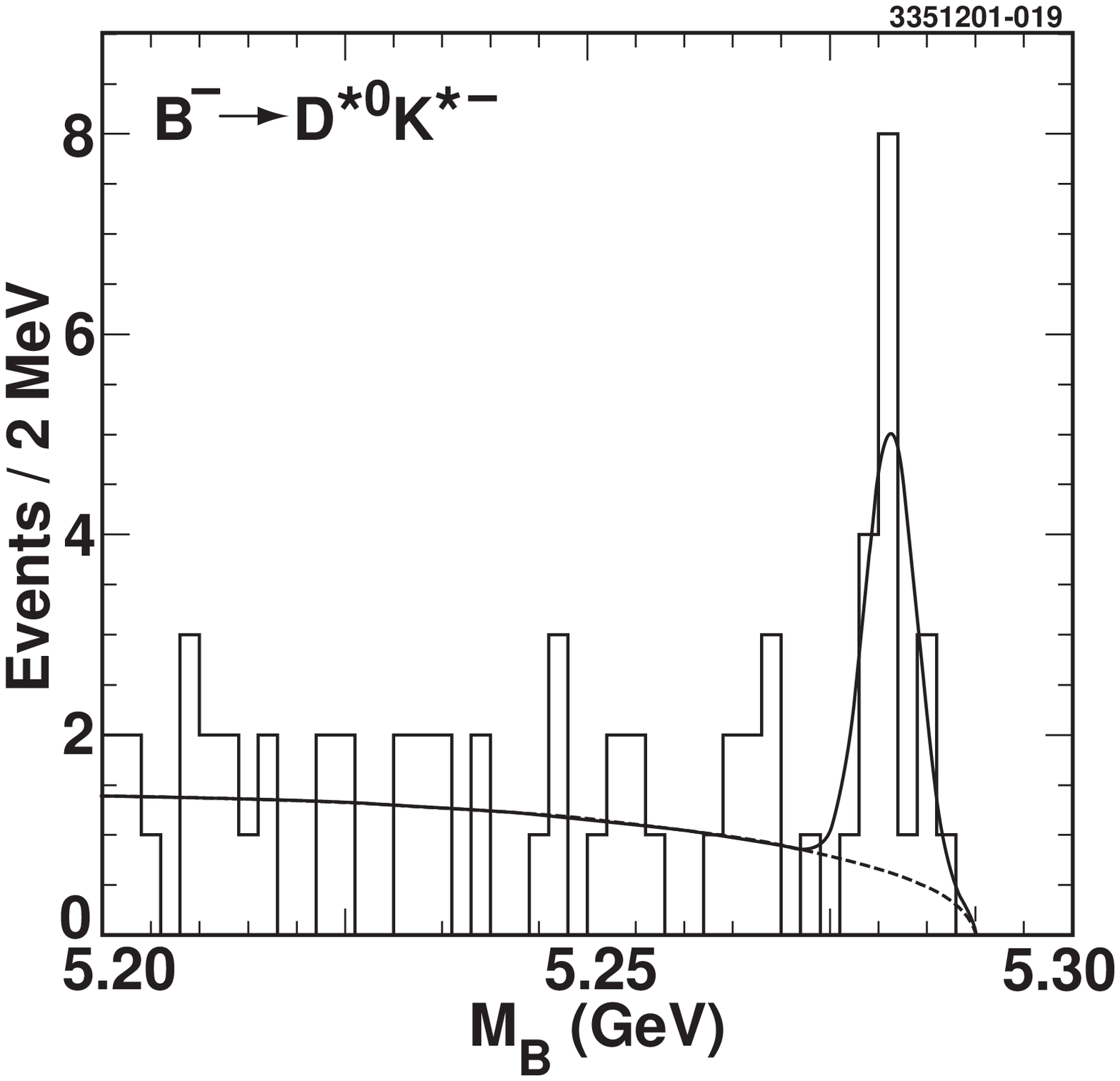}}}
\caption{The  beam-constrained mass distributions for 
         $\bar B\to D^{(*)} K^{*-}$ decays 
         with all sub-modes combined. 
         The histograms show the data, the solid lines represent
         the overall fit to the data, and the dashed lines represent
         the fitted backgrounds under the peaks.}  
\label{fig01}
\end{figure}

We combine all the sub-modes for each $\bar B \to D^{(*)}K^{*-}$ decay 
and fit 
the beam-constrained mass distributions using a binned maximum likelihood
method with the mass resolution fixed at the value determined from
Monte Carlo  simulation. Studies of
similar $\bar B\to D^{(*)}\pi^-$ modes show that the mass resolutions 
in the Monte Carlo samples and in the data are in good agreement. 
 The beam-constrained mass 
distribution is modeled by a Gaussian plus
the phenomenological background function \cite{ARGUS}. 
The results for the $\bar B \to D^{(*)}K^{*-}$ modes 
are shown in Fig.~\ref{fig01} and summarized in Table ~\ref{table1}.
Significant excesses of events are observed in the signal regions, 
compatible with our typical $M_B$ and $\Delta E$ resolutions. 
These excesses are consistent with coming from 
$\bar{B} \to D^{(*)}K^{*-}$ decays.

We use wrong-sign ($D^{(*)}K^{*+}$) 
combinations from the on-resonance sample 
and right-sign combinations from the continuum sample as cross-checks.
For the wrong sign analysis, instead of searching for 
$D^{(*)}K^{*-}$ combinations where 
$K^{*-} \to K^{0}_{S} \pi^{-}$, we search for $D^{(*)}K^{*+}$ combinations 
where $K^{*+} \to K^{0}_{S} \pi^{+}$.
 The same selection cuts are applied for the wrong-sign candidates 
as for the right-sign ones. 
No enhancement is observed in the signal region for these wrong-sign 
combinations. 
We also apply the same selection criteria to 
the right-sign combinations in the continuum sample, 
and no enhancement is observed in the signal region.

\begin{table}[h]
\caption{Results for the $\bar{B}$ $\to$ $D^{(*)} K^{*-}$ decay modes
         with statistical errors only. 
         For $\bar{B}$ $\to$ $D^{*} K^{*-}$, the
         two efficiencies correspond to the 00 and 11 helicity states, 
         respectively. The statistical significance of the 
         overall signal for each mode is given in 
         parentheses.}
 \begin{center}
 \begin{tabular}{ l c c c } 
         {       Decay Mode}      
    &    {       Efficiency(\%)}  
    &    {       Yield}                              
    &    {       $\cal{B}$($\times10^{-4}$)}         \\ \hline
         {       $B^{-} \to D^{0}(K^{-}\pi^{+}) K^{*-}$}
    &    {        18.7}  
    &    {        $9.8 \pm 3.7$}                                       
    &    {        $6.2 \pm 2.3$}            \\
         {        $B^{-} \to D^{0}(K^{-}\pi^{+}\pi^{0}) K^{*-}$}
    &    {        5.4}  
    &    {        $7.9 \pm 4.4$}                                       
    &    {        $4.9 \pm 2.7$}       \\ 
         {        $B^{-} \to D^{0}(K^{-}\pi^{+}\pi^{+}\pi^{-}) K^{*-}$}
    &    {        10.7}  
    &    {        $11.9 \pm 5.6$}                                       
    &    {        $6.7 \pm 3.2$}        \\ \hline
         {         $B^{-} \to D^{0} K^{*-}$}
    &    {             }  
    &    {        $30.5 \pm 8.3~(6.0\sigma)$}
    &    {         $6.1 \pm 1.6$}       \\ \hline
         {        $\bar{B^{0}} \to D^{+}(K^{-}\pi^{+}\pi^{+}) K^{*-}$}
    &    {        15.5}  
    &    {        $11.5 \pm 4.7~(4.2\sigma)$}
    &    {         $3.7 \pm 1.5$}        \\ \hline
         {        $\bar{B^{0}} \to D^{*+}(D^{0}(K^{-}\pi^{+})\pi^{+}_{s})
                        K^{*-}$}
    &    {        14.9, 17.1}  
    &    {        $0.8 \pm 1.1$}                                       
    &    {        $^{0.9~\pm~1.3}_{0.8~\pm~1.1} $} \\ 
         {        $\bar{B^{0}} \to D^{*+}(D^{0}(K^{-}\pi^{+}\pi^{0}) 
                        \pi^{+}_{s}) K^{*-}$}
    &    {        4.7, 5.5}  
    &    {        $4.3 \pm 2.3$}                                       
    &    {        $^{4.5~\pm~2.4}_{3.8~\pm~2.0} $} \\ 
         {        $\bar{B^{0}} \to D^{*+}(D^{0}(K^{-}\pi^{+}\pi^{+}
                        \pi^{-}) \pi^{+}_{s}) K^{*-}$}
    &    {        6.5, 8.1}  
    &    {        $5.5 \pm 2.5$}                                       
    &    {        $^{7.5~\pm~3.4}_{6.1~\pm~2.8} $} \\ \hline
         {         $\bar{B^{0}} \to D^{*+} K^{*-}$}
    &    {              }  
    &    {         $10.5 \pm 3.5~(5.4\sigma)$}                           
    &    {         $^{4.1~\pm~1.4}_{3.5~\pm~1.2} $} \\ \hline
         {       $B^{-} \to D^{*0}(D^{0}(K^{-}\pi^{+})\pi^{0}) 
                        K^{*-}$}
    &    {        6.7, 8.0}  
    &    {        $-0.12 \pm 0.07$}                                       
    &    {        $^{-0.3~\pm~0.2}_{-0.3~\pm~0.2}$ } \\ 
         {        $B^{-} \to D^{*0}(D^{0}(K^{-}\pi^{+}\pi^{0})\pi^{0}) 
                        K^{*-}$}
    &    {        2.3, 2.6}  
    &    {        $3.6 \pm 2.0$}                                       
    &    {        $^{8.4~\pm~4.7}_{7.5~\pm~4.1} $} \\ 
         {        $B^{-} \to D^{*0}(D^{0}(K^{-}\pi^{+}\pi^{+}\pi^{-}) 
                        \pi^{0}) K^{*-}$}
    &    {        3.5, 4.4}  
    &    {        $0.8 \pm 1.5$}                                       
    &    {        $^{2.3~\pm~4.3}_{1.8~\pm~3.4} $} \\ 
         {        $B^{-} \to D^{*0}(D^{0}(K^{-}\pi^{+})\gamma) 
                        K^{*-}$}
    &    {        6.0, 6.9}  
    &    {        $2.9 \pm 1.7$}                                       
    &    {        $^{15.0~\pm~8.8}_{13.0~\pm~7.6} $} \\ 
         {        $B^{-} \to D^{*0}(D^{0}(K^{-}\pi^{+}\pi^{0})\gamma) 
                        K^{*-}$}
    &    {        1.9, 2.0}  
    &    {        $2.7 \pm 1.7$}                                       
    &    {        $^{12.4~\pm~7.8}_{11.8~\pm~7.4} $}  \\ 
         {        $B^{-} \to D^{*0}(D^{0}(K^{-}\pi^{+}\pi^{+}\pi^{-})
                        \gamma) K^{*-}$}
    &    {        3.5, 3.7}  
    &    {        $6.0 \pm 2.7$}                                       
    &    {        $^{27.2~\pm~12.3}_{25.7~\pm~11.6} $} \\ \hline
         {         $B^{-} \to D^{*0} K^{*-}$}
    &    {             }  
    &    {         $14.6 \pm 4.3~(6.2\sigma)$}                            
    &    {         $^{8.3~\pm~2.4}_{7.2~\pm~2.1} $} 
\end{tabular}
\end{center}
\label{table1}
\end{table}

To estimate the detection efficiencies, we generated 
$\bar B\to D^{(*)}K^{*-}$ Monte Carlo events 
and  simulated the CLEO detector 
response with GEANT \cite{GEANT}.  Simulated events
for the CLEO II and II.V configurations were processed in the 
same manner as the data. The resulting detection 
efficiencies, which are listed in Table~\ref{table1},   
do not include the $D^{(*)}$ or $K^{*-}$ decay branching fractions~\cite{pdg}.
For the $\bar B \to D^{*} K^{*-}$ decays, as we do not know the polarization of
the two vector mesons, we generated both 00 and 11 helicity states 
for our Monte Carlo signal events to calculate  the efficiencies
as shown in Tables~\ref{table1}.


Systematic errors from event selection include uncertainties in 
$dE/dx$, 
the Dalitz weight cut for $D^0\to K^-\pi^+\pi^0$ and Monte Carlo statistics. 
To estimate the effects of  the $\pi^{0}$ veto and 
the $E_{\gamma}$ $>$ 100 MeV requirement for the photon candidates in
$D^{*0} \to D^{0}\gamma$ decays, we change the $\pi^{0}$ veto
interval by 1$\sigma$ and the energy cut value by $\pm$ 10 MeV 
respectively,
and take the difference in the efficiency-corrected yields as the systematic 
error. 
The cross-feed rates between
$\bar B \to D^{*0} K^{*-}$, where $D^{*0} \to D^{0}\pi^{0}$ and
$D^{*0} \to D^{0}\gamma$, are less than 1.5\%.
We also use tighter and looser requirements of: $1\sigma/3\sigma$ 
for the $D^{(*)}$
mass cut, 50 MeV/100 MeV for the $K^{*-}$ mass requirement, 
and 0.8/0.9 for the cosine of the sphericity angle to
obtain new signal yields, efficiencies and branching ratios. 
Since the data were collected with two different detector configurations,
we have also calculated the branching ratios separately for the two sets of
data. If the difference in the branching ratios
between the two data sets is larger than the one from
varying the selection cuts, this is taken as the systematic error.
The variation of cuts contributes the major part of the total  
systematic error. 
Systematic errors also 
include uncertainties from  track, shower,
 $K_S^0$ and $\pi^0$ reconstruction efficiencies and uncertainties
from the number of $B\bar B$ pairs,
the ratio of $B^0\bar B^0$ to $B\bar B$ in $\Upsilon(4S)$
decays~\cite{bbratio} and the
$D^{(*)}$ and $K^{*-}$ decay branching fractions~\cite{pdg}. 

To study the systematic errors from the $\bar{B} \to D^{(*)} K^{*-}$ 
fitting procedure, we fit the mass distributions using a 
background shape obtained from 
the continuum data for each decay mode. The differences with 
Table~\ref{table1}, which are less than 3\%,   
 are taken as the systematic errors of the signal yields.
The systematic errors due to uncertainties of the 
beam-constrained mass resolutions determined from Monte Carlo 
simulation are less than 3\% from studies of similar 
$\bar B\to D^{(*)}\pi^-$ decays.  

The efficiency for observing $\bar B\to D^{*}K^{*-}$ 
decays depends on the unknown polarization of the final state. Assuming
both final state mesons are in a helicity 0 state, we find 
${\cal B}(\bar{B^{0}}\to D^{*+}K^{*-})=(4.1~\pm~1.4~\pm~0.8)\times 10^{-4}$ and 
${\cal B}(B^{-} \to D^{*0} K^{*-} )=(8.3~\pm~2.4~\pm~2.6)\times 10^{-4}$. 
Should the dynamics of the decay process force the final state mesons into
$\pm1$ helicity states, we find 
${\cal B}(\bar{B^{0}}\to D^{*+}K^{*-})=(3.5~\pm~1.2~\pm~0.7)\times 10^{-4}$
and ${\cal B}(B^{-} \to D^{*0} K^{*-} )=(7.2~\pm~2.1~\pm~2.4)\times 10^{-4}$.
The errors 
are statistical and systematic.   
Assuming an unpolarizaed final state, 
we take the equally-weighted 
average of the two branching ratios, corresponding to the 
00 and 11 helicity states from Table~\ref{table1}, as the  
central value, and the differences 
as a  systematic error. 
These differences are small
compared to the statistical errors. 


We have observed significant signals in the exclusive 
$\bar B \to D^{(*)} K^{*-}$ decay decays. 
The measured branching ratios are:

$${\cal B}(B^{-} \to D^{0} K^{*-} ) =     (6.1~\pm~1.6~\pm~1.7)\times 10^{-4},$$
$${\cal B}(\bar{B^{0}}\to D^{+}K^{*-}) =  (3.7~\pm~1.5~\pm~1.0)\times 10^{-4},$$
$${\cal B}(\bar{B^{0}}\to D^{*+}K^{*-}) = (3.8~\pm~1.3~\pm~0.8)\times 10^{-4},$$
$${\cal B}(B^{-} \to D^{*0} K^{*-} ) = (7.7~\pm~2.2~\pm~2.6)\times 10^{-4}.$$   
The errors shown are
statistical and systematic, respectively.

With the assumption of the validity of factorization,
$SU(3)$ flavor symmetry relates the Cabibbo-suppressed decays
$\bar B\to D^{(*)}K^{*-}$ to the Cabibbo-favored ones
$\bar B\to D^{(*)}\rho^-$
by $\frac{{\cal B}(\bar B\to D^{(*)}K^{*-})}
{{\cal B}(\bar B\to D^{(*)}\rho^-)}\simeq
\left|\frac{V_{us}}{V_{ud}}\right|^2\left(\frac{f_{K^{*-}}}{f_{\rho^-}}
\right)^2
\simeq5\%$ in the tree level approximation. Here, $V_{us}$ and $V_{ud}$
are CKM elements,
and $f_{K^{*-}}$ and $f_{\rho^-}$ are the meson decay constants
which can be derived from $\tau^-\to K^{*-}\nu_{\tau}$
and  $\tau^-\to \rho^-\nu_{\tau}$ decays. 
For the above decay modes, the measured ratios 
$\frac{{\cal B}(\bar B\to D^{(*)}K^{*-})}
{{\cal B}(\bar B\to D^{(*)}\rho^-)}$ are (4.6 $\pm$ 1.8)\%,
(4.7 $\pm$ 2.4)\%, (5.6 $\pm$ 3.6)\% and (5.0 $\pm$ 2.4)\% 
respectively, which are consistent with 
naive theoretical predictions within the errors~\cite{naive}. The branching
ratios for $\bar B\to D^{(*)}\rho^-$ decays are taken from PDG 
2000~\cite{pdg}. Our results ${\cal B}(\bar{B^{0}}\to D^{(*)+}K^{*-})$
are in good agreement with the theoretic predictions within the errors
\cite{neubert}. 
The rates for $\bar B\to D^{(*)}K^{*-}$
 suggest that these decays  could provide  a  measurement of the 
angle $\gamma$ in the unitary triangle using the color-allowed decays
in the near future~\cite{color}. 

\acknowledgements
We gratefully acknowledge the effort of the CESR staff in providing us with
excellent luminosity and running conditions.
This work was supported by 
the National Science Foundation,
the U.S. Department of Energy,
the Research Corporation,
and the Texas Advanced Research Program.


\begin{thebibliography}{99}

\bibitem{ckm} M. Kobayashi and T. Maskawa, Prog. Theor. Phys. {\bf 49},
              652 (1973).

\bibitem{babarbook}
    P. F. Harrison and H. R. Quinn,
    The BaBar Physics Book (SLAC-R-0504, 1998).

\bibitem{gronau}
    M. Gronau and D. Wyler, Phys. Lett. {\bf B265}, 172 (1991);
    I. Dunietz, Phys. Lett. {\bf B270}, 75 (1991);
    D. Atwood, G. Eilam, M. Gronau and A. Soni, 
               Phys. Lett. {\bf B341}, 372 (1995); 
    D. Atwood, I. Dunietz and A. Soni, Phys. Rev. Lett. {\bf 78}, 3257 (1997).

\bibitem{cleodk}
 CLEO Collaboration,    M. Athanas {\it et al.}, 
    Phys. Rev. Lett. {\bf 80}, 5493 (1998).

\bibitem{BELLE} BELLE Collaboration, K. Abe {\it et al.}, 
                  Phys. Rev. Lett. {\bf 87}, 111801 (2001). 

\bibitem{CLEOII}  CLEO Collaboration, Y. Kubota {\it et al.}, Nucl. Instrum.
                  Methods  A {\bf 320}, 66 (1992).


\bibitem{CLEOII.V} T.~Hill, Nucl. Instrum.
                  Methods  A {\bf 418}, 32 (1998).


\bibitem{bbratio} 
CLEO Collaboration, J. Alexander {\it et al.}, Phys. Rev. Lett. {\bf 86},
2737, (2000). 
\bibitem{pdg} Particle Data Group, D. E. Groom {\it et al.},  
      Eur. Phys. J. {\bf C15}, 1 (2000).

\bibitem{Dalitz} CLEO Collaboration, S. Kopp {\it et al.}, 
Phys. Rev. D {\bf 63}, 092001 (2001).  

\bibitem{sphericity} S.~L.~Wu, Phys.\ Rep.\ C {\bf 107}, 59 (1984).

\bibitem{ARGUS}ARGUS Collaboration, H. Albrecht {\it et al.}, Z. Phys. 
C {\bf 48}, 543 (1990).
            

\bibitem{GEANT} R. Burn {\it et al.}, GEANT3.14, 
CERN DD/EE/84-1 (unpublished). 


\bibitem{naive}
Similar ratios  are
$\frac{{\cal B}(\bar B\to D^{(*)}K^{-})} {{\cal B}(\bar B\to D^{(*)}\pi^-)}
\simeq \left|\frac{V_{us}}{V_{ud}}\right|^2\left(\frac{f_{K}}{f_{\pi}}\right)^2
\simeq7\%$.  
The results of 
$\frac{{\cal B}(\bar B\to D^{(*)}K^{-})} {{\cal B}(\bar B\to D^{(*)}\pi^-)}$ 
from BELLE are consistent with the naive model 
calculation, see Ref.~\cite{BELLE}. 
\bibitem{neubert}M. Beneke, G. Buchalla, M. Neubert and C. T. Sachrajda, 
Nucl. Phys. {\bf B591}, 313 (2000). 
\bibitem{color}M. Gronau, Phys. Rev. D {\bf 58}, 037301 (1998);
J. H. Jang and P. Ko, Phys. Rev. D {\bf 58}, 111302 (1998).   



\end{thebibliography}
\end{document}